\documentclass[apjl]{emulateapj}
\usepackage[]{graphicx}

\newcommand{\kms}{\hbox{${\rm km\;s}^{-1}$}}

\slugcomment{Accepted by ApJ Letters}

\shorttitle{Anti-Truncations in Early-Type Disks}
\shortauthors{Erwin et al.}

\begin{document}

\title{Anti-Truncation of Disks in Early-Type Barred Galaxies}

\author{Peter Erwin, John E. Beckman, and Michael Pohlen}
\affil{Instituto de Astrof\'{\i}sica de Canarias, C/ Via L\'{a}ctea s/n, 
38200 La Laguna, Tenerife, Spain}
\email{erwin@iac.es, jeb@iac.es, pohlen@iac.es}

\begin{abstract} 

The disks of spiral galaxies are commonly thought to be truncated: the
radial surface brightness profile steepens sharply beyond a certain
radius (3--5 inner-disk scale lengths).  Here we present the radial
brightness profiles of a number of barred S0--Sb galaxies with the
opposite behavior: their outer profiles are distinctly shallower in
slope than the main disk profile.  We term these ``anti-truncations'';
they are found in at least 25\% of a larger sample of barred S0--Sb
galaxies.  There are two distinct types of anti-truncations.  About
one-third show a fairly gradual transition and outer isophotes which
are progressively rounder than the main disk isophotes, suggestive of
a disk embedded within a more spheroidal outer zone --- either the
outer extent of the bulge or a separate stellar halo.  But the
majority of the profiles have rather sharp surface-brightness
transitions to the shallower, outer exponential profile and,
crucially, outer isophotes which are not significantly rounder than
the main disk; in the Sab--Sb galaxies, the outer isophotes include
visible spiral arms.  This suggests that the outer light is still part
of the disk.  A subset of these profiles are in galaxies with
asymmetric outer isophotes (lopsided or one-armed spirals), suggesting
that interactions may be responsible for at least some of the disklike
anti-truncations.

\end{abstract}


\section{Introduction} 

Truncations in the stellar population at the edges of disk galaxies
are thought to be a common morphological feature \citep[for a recent
review, see][]{pohlen04b}.  \nocite{vdk79}Van der Kruit (1979) and
\nocite{vdk81a,vdk81b}van der Kruit \& Searle (1981a, 1981b), using
photographic images of mostly late-type, edge-on spirals, found that
the exponential decline in surface brightness in the inner disks
seemed to change quite abruptly to a steeper decline at the edges of
the galaxies.  In principle, the faint surface brightness of the outer
disk region (face-on $\mu_{B} \gtrsim 25$) makes truncations easier to
detect in edge-on disks, where line-of-sight integration increases the
observed surface brightness.  However, their very edge-on nature makes
it difficult to determine if detected truncations are truly
\textit{radial} in nature, as opposed to azimuthal variations due to,
e.g., strong spiral arms.  The difficulty of clearly identifying disk
features (spirals, bars, rings, etc.)  in edge-on galaxies also makes
it hard to see whether truncations might be related to specific
morphological and dynamical features in the disk.

\citet{pohlen02}, using very deep exposures of three face-on Sbc--Sc
galaxies, showed that truncations can be clearly identified in face-on
disks, and that the decline in surface brightness outside the
truncation break is best described by an exponential, steeper by about
a factor of two than that for the inner disk.  Figure~\ref{fig:one}
shows an example of such a ``classical'' truncation.  Despite the
apparent prevalence of truncations at 3--5 scale lengths in edge-on
galaxies \citep[principally late-type spirals;
e.g.,][]{pohlen01,degrijs01,kregel02}, there \textit{are} face-on or
moderately inclined galaxies where the disk's surface brightness
profile can be traced as a single exponential to larger radii without
truncation \citep[e.g.,][]{barton97,weiner01,erwin05};
Figure~\ref{fig:one} shows an example.

In this Letter we draw attention to a significant subset (at least
$\sim 25$\%) of barred S0--Sb galaxies which exhibit a kind of
``anti-truncation'': the surface brightness profile becomes
\textit{shallower} at large radii, so that there is an excess of light
above the outward projection of the (inner) exponential profile.
While some of these profiles can be explained as light from a round
outer envelope or halo (perhaps the outer extent of the bulge)
dominating at larger radii, the majority cannot.  Instead, they appear
to be continuations of the disk with shallower but still exponential
profiles.  Some of these galaxies show strong asymmetries (e.g.,
lopsided or one-armed spirals) in the outer disk, and so may represent
the result of recent interactions.

\section{Sample Selection and Data} 

The overall sample on which this study is based is described in more
detail in \citet{erwin05-bars} and \nocite{erwin05}Erwin et al.\
(2005).  Briefly, it consists of all northern ($\delta > -10\arcdeg$),
barred S0--Sb galaxies from the Uppsala Galaxy Catalog \citep{ugc}
with $D_{25} \geq 2.0\arcmin$, axis ratio $a/b \leq 2.0$ ($i \lesssim
60\arcdeg$), and heliocentric $V \leq 2000$ \kms{} \citep[measurements
and classifications from][hereafter RC3]{rc3}.  Because there is
evidence for inconsistency in Hubble types when spirals in the Virgo
Cluster are compared with field spirals \citep[e.g., Virgo Sa galaxies
tend to resemble field Sc galaxies;][]{koopman98}, we excluded Virgo
Cluster galaxies.  (We did, however, include S0 galaxies from the
Virgo Cluster, since there is no evidence for inconsistent
classification of these galaxies.)  After eliminating a few galaxies
without bars \citep[see][]{erwin05-bars}, we had a total of 65
galaxies.

The sample is complete within its selection criteria, but the diameter
limit produces a bias in favor of luminous and high-surface-brightness
galaxies.  It is obviously biased against \textit{unbarred} galaxies;
but since we included both strong and weakly barred galaxies (SB and
SAB classifications from \nocite{rc3}RC3), the result is
representative of the majority of local, early-type disk galaxies.

We obtained azimuthally averaged, $R$-band surface brightness profiles
of all the objects.  Details of the observations, calibrations, and
reduction procedures are given in \nocite{erwin05}Erwin et al.\
(2005).  Particular attention was paid to ensuring flat and accurately
subtracted sky backgrounds, with the halos of bright stars and
neighboring galaxies carefully masked out.  The orientation of the
outer disk (position angle and projected ellpiticity) was derived from
free-ellipse fits to the images, or from kinematic studies in the
literature \citep[see][]{erwin-sparke03,erwin05-bars}.  The final
profiles are from logarithmically spaced, concentric ellipses with
constant position angle and ellipticity, matching that of the outer
disk.

\section{Anti-Truncations in Surface Brightness Profiles}

\begin{figure}
\begin{center}
\includegraphics[scale=0.5]{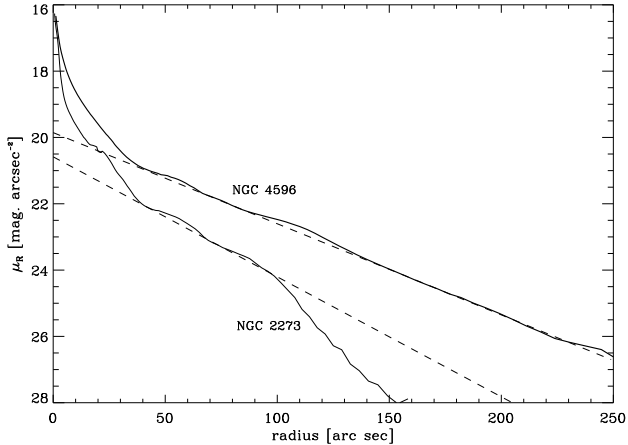}
\end{center}

\caption{Truncated and untruncated surface-brightness profiles of
barred galaxies: azimuthally averaged surface-brightness profiles for
NGC~4596 (SB0) and NGC~2273 (SBa), along with exponential fits (dashed
lines).  NGC~4596 shows a Type~I exponential profile extending to at
least 5.7 scale lengths with no sign of a truncation; NGC~2273 shows a
truncation at $r \sim 3.3$ scale lengths.\label{fig:one}}

\end{figure}

Figure~\ref{fig:one} shows surface brightness profiles for two
moderately inclined galaxies which exhibit ``conventional'' behavior:
NGC~2273 shows a classical truncation at 3.3 scale lengths of the
inner ($r \lesssim 95\arcsec$) disk, while NGC~4596 shows a single
exponential profile which extends to at least $> 6$ scale lengths with
no hint of a truncation.  The majority of galaxies in our sample have
similar profiles, with untruncated profiles being far more common (see
\nocite{erwin05}Erwin et al.\ 2005 for the complete set of profiles).

\begin{figure}
\begin{center}
\includegraphics[scale=1.0]{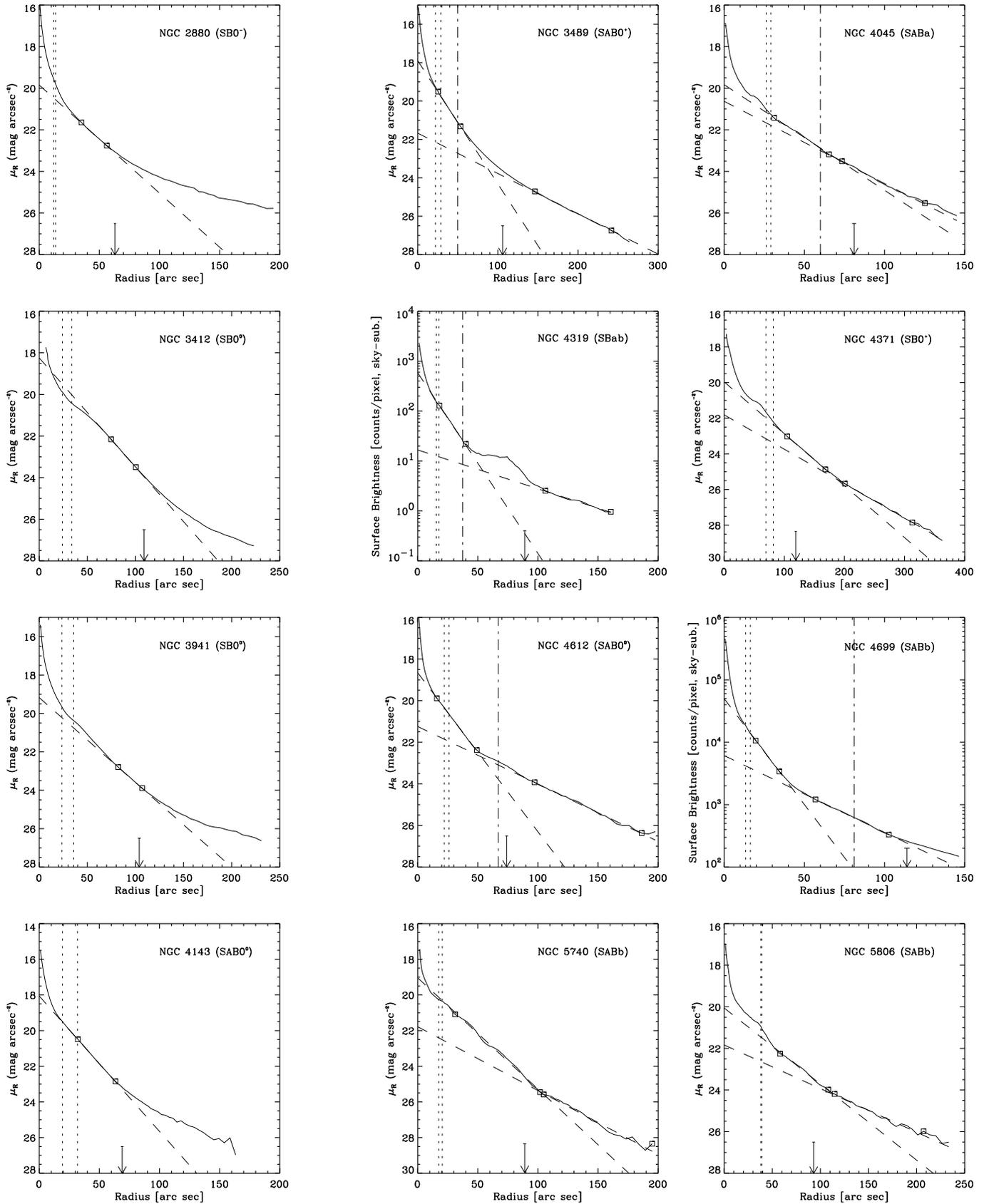}
\end{center}

\caption{Type III (``anti-truncation'') $R$-band surface-brightness
profiles.  \textit{Left column}: Outer excess light is associated with
halo/spheroid.  \textit{Center and right columns}: Outer excess light
is probably part of the disk.  Vertical dashed lines mark lower- and
upper-limit measurements of the bar semi-major axis (deprojected);
vertical dot-dashed lines indicate approximate size of outer rings, if
any; and slanted dashed lines are exponential fits to regions of the
profile delimited by the small squares.  The arrows indicate
$R_{25}$ (one-half of $D_{25}$).  No photometric calibration was
possible for NGC~4319 or NGC~4699.\label{fig:big}}

\end{figure}

At least 25\% of the galaxies have profiles with qualitatively
different behavior: there is an excess of light at large radii, above
the projected exponential disk profile.  The transition happens at
surface brightness levels $\mu_{R} \sim 22.6$--25.6 mag arcsec$^{-2}$
(mean = $24.1 \pm 1.0$; for comparison, the truncations in
\nocite{pohlen02}Pohlen et al.\ 2002 are at a mean $\mu_{R} = 24.6$).
Figure~\ref{fig:big} presents profiles for 12 of the 16 galaxies which
unambiguously show this behavior (there are other galaxies which may
have excess light at large radii, but S/N and sky-subtraction
uncertainties make it more difficult to be certain).  Because this is
the opposite of a classical truncation --- and happens at similar
radii (3.2--6.0 inner scale lengths) --- we dub them
``anti-truncations''; by extending \nocite{freeman70}Freeman's (1970)
classification scheme we can also refer to them as ``Type III''
profiles.

\begin{figure}
\begin{center}
\includegraphics[scale=0.95]{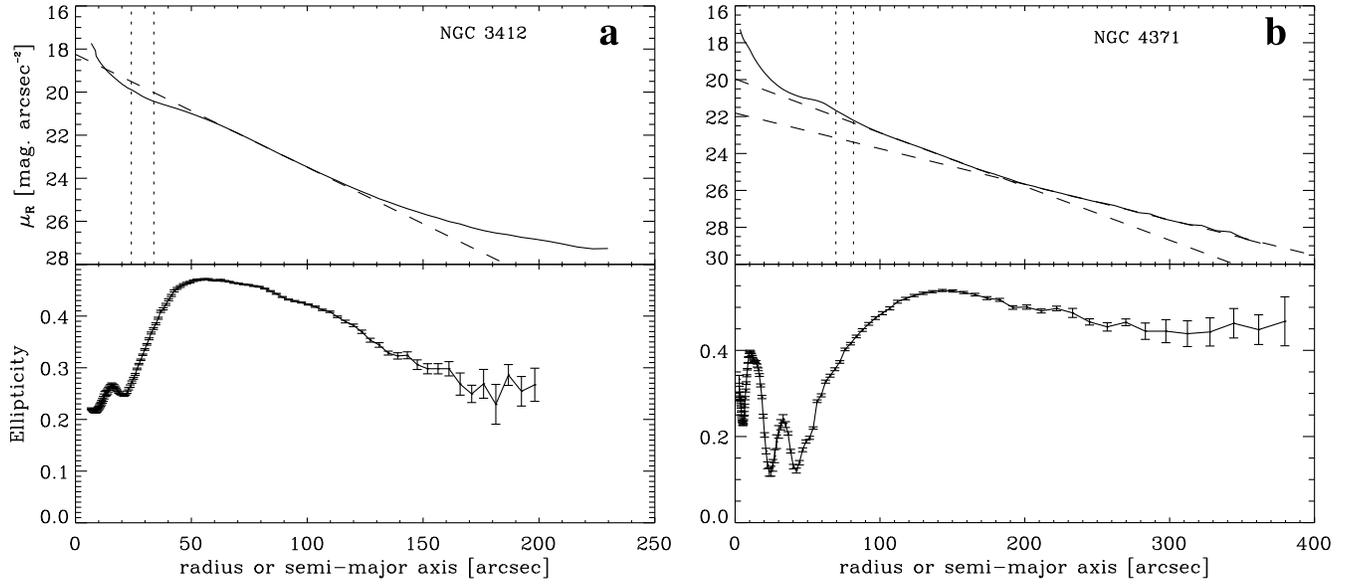}
\end{center}

\caption{Azimuthally averaged surface-brightness profiles (top) and
isophotal ellipticity profiles (bottom) for two galaxies with Type III
profiles.  \textbf{a.} NGC~3412 shows a clear ``disk + spheroid''
profile: the outer excess light ($r \gtrsim 120\arcsec$) is associated
with isophotes which become progressively rounder at larger radii,
consistent with the idea of a (projected) elliptical disk embedded
within a rounder spheroid, e.g., a halo or outer bulge.  \textbf{b.}
NGC~4371 has outer isophotes ($r \gtrsim 190\arcsec$) with
approximately constant ellipticity, roughly the same as the inner
isophotes, so the excess light at $r \gtrsim 190\arcsec$ is probably
part of the disk.\label{fig:ell}}

\end{figure}

Because these are early-type galaxies (many of them S0), an obvious
interpretation would be that the excess light at large radii is not
from the \textit{disk}, but rather from the luminous bulge or halo.
This appears to be true for about one-third of the galaxies with
anti-truncations (left-hand column of Figure~\ref{fig:big}) --- all of
them S0 --- which have two characteristics in common.  First, the
inflection in the profiles is smooth and curved, which suggests that
the outer light is from a separate component, added to the inner light
from the disk.  Second, the isophotes are elliptical in the inner
region and become progressively \textit{rounder} at larger radii
(Figure~\ref{fig:ell}a); this is clear evidence for an inclined disk
embedded in a more spheroidal outer component.

But the majority of the galaxies (center and right-hand columns of
Figure~\ref{fig:big}) have outer isophotes which are \textit{not}
significantly rounder than the inner isophotes
(Figure~\ref{fig:ell}b).  Morphologically at least, the excess outer
light is still part of the disk.  In several of the Sab and Sb
galaxies (e.g., NGC~4319, 4699, 5740, 5806), symmetric or
asymmetric spiral arms are clearly visible in the region outside the
transition, which again argues for the excess outer light being part
of the disk.  Finally, in most of these profiles the transition is
quite sharp, so we are probably not seeing the addition of light from
two overlapping components, but rather a single disk with a fairly
abrupt change in its density profile.  The outer part of the profile
is usually quite exponential; we show exponential fits to the outer
profiles in Figure~\ref{fig:big}.

Could we be seeing thick disks, which are exponential but with larger
scale lengths than the brighter thin disks?  The combined profile of a
thin disk and a thick disk should be the sum of two exponentials;
observations of edge-on galaxies show that scale length ratios (thick
to thin) are almost always less than 2.0 \citep{pohlen04a}.  For some
of our galaxies, particularly those with smooth transitions between
the inner and outer regions, the profile is indeed well fit by a
double-exponential (e.g., NGC~3489, 3941, and 4143).  However, the
resulting scale length ratios (outer to inner, or ``thick'' to
``thin'') range from 2.1 to 6.3, with a mean of 3.8, which is clearly
outside the range of known thin/thick-disk systems.  Furthermore, for
the galaxies with the sharpest transitions (e.g., NGC~3982, 4371,
4691, and 5740), the best-fitting double-exponential produces a
distinct excess above the observed profile in the transition region.
We conclude that the outer excesses we see are probably \textit{not}
thick disks.

So what's going on?  One clue may be the presence of
lopsided/asymmetric spiral arms in the outer regions of some of the
``disky'' galaxies.  NGC~4319 has a dramatic, one-armed outer spiral
(responsible for the excess light between 50\arcsec{} and 100\arcsec{}
in the profile; Figure~\ref{fig:big}) which may be connected to a
faint structure further out resembling a tidal tail.  NGC~5740 and
NGC~5806 have weaker, lopsided spiral distortions in their outer
isophotes.  So one possibility is that interactions are responsible in
some fashion for the outer excess light in the disks.

In two galaxies (NGC~4045 and NGC~4612), there are outer rings
coincident with the transition between inner and outer slopes; in
three more (NGC~4371, NGC~4699, and NGC~5806), the transition happens
at $\sim 2$--2.5 times the bar radius, which is the typical size of
outer rings \citep[e.g,][]{buta93}.  This suggests that the transition
might be happening at or near the bar's outer Lindblad resonance.

We note that in some galaxies (e.g., NGC~4612 and NGC~4699), the
transition to the outer profile happens at a relatively small radius
and high surface brightness level; since the outer disk is easily
detected in these galaxies, it might be tempting to classify the inner
exponential slope as the galaxy's ``bulge'' \citep[the $r \lesssim
40\arcsec$ region of NGC~4699 is in fact referred to this way in
][]{carnegie}.  But these inner regions are where the bars (and, in
NGC~4699, numerous spiral arms and dust lanes) are found and are thus
\textit{not} classical (kinematically hot) bulges; smaller central
excesses (e.g., $r \lesssim 10\arcsec$ for NGC~4699) may be better
candidates for the bulge proper.  The inner disk regions of these two
galaxies, at least, can thus be viewed as ``pseudobulges''
\citep[e.g.,][]{kormendy04}, albeit much larger than those
pseudobulges thought to arise from bar-driven inflow, which are
necessarily smaller than the bars.

These surface-brightness profiles with outer excesses, which we have
termed anti-truncations or Type~III profiles, are found in $\sim 25$\%
of our sample; this should be regarded as a lower limit, since not all
of our images reach the same limiting surface brightness and we could
be missing similar features at fainter surface-brightness levels.
Even if we exclude those galaxies whose isophote shapes indicate we
are probably seeing luminous halos or outer bulges (left-hand column
of Figure~\ref{fig:big}; 8\% of the full sample, or 20\% of the S0
galaxies), at least 17\% of the galaxies appear to have disks with
anti-truncation behavior.  Disks with anti-truncations thus appear to
be \textit{more} common than disks with classical truncations
($\lesssim 12$\% of the sample; see \nocite{erwin05}Erwin et al.\
2005), at least in barred, S0--Sb disk galaxies.

\acknowledgements

We would like to thank the anonymous referee for a careful reading,
and especially for suggesting we take a closer look at the issue of
thick disks.  This research was partly supported by grant
AYA2004-08251-CO2-01 from the Spanish Ministerio de Ciencia y
Educaci\'{o}n.


\end{document}